\documentclass[amsmath,amssymb,aps,prl,twocolumn,floatfix]{revtex4-2}

\usepackage{amssymb}
\usepackage{graphicx}
\usepackage{dcolumn}
\usepackage{bm}
\usepackage{amsmath}
\usepackage{mathrsfs}
\usepackage{textcomp}
\usepackage[unicode=true,bookmarks=true,bookmarksnumbered=false,bookmarksopen=false,breaklinks=false,pdfborder={0 0 1},backref=false,colorlinks=true]{hyperref}

\hypersetup{linkcolor=magenta,urlcolor=blue,citecolor=blue,pdfstartview={FitH},unicode=true}

\def\be{\begin{equation}}
\def\ee{\end{equation}}
\def\bea{\begin{eqnarray}}
\def\eea{\end{eqnarray}}

\begin{document}

\preprint{APS/123-QED}

\title{Interference and short-range correlation in fermionic Hubbard gases}

\author{Yan-Song Zhu$^{1,2}$}
\author{Hou-Ji Shao$^{1,2}$}
\author{Yu-Xuan Wang$^{1,2}$}
\author{De-Zhi Zhu$^{1,2}$}
\author{Hao-Nan Sun$^{1,2}$}
\author{Si-Yuan Chen$^{1,2}$}
\author{Chi Zhang$^{1,2}$}
\author{Xing-Can Yao$^{1,2,3}$}
\thanks {yaoxing@ustc.edu.cn}
\author{Yu-Ao Chen$^{1,2,3,4}$}
\thanks{yuaochen@ustc.edu.cn}
\author{Jian-Wei Pan$^{1,2,3}$}
\thanks{pan@ustc.edu.cn}

\affiliation{$1$Hefei National Research Center for Physical Sciences at the Microscale and School of Physical Sciences, University of Science and Technology of China, Hefei 230026, China}
\affiliation{$2$Shanghai Research Center for Quantum Science and CAS Center for Excellence in Quantum Information and Quantum Physics, University of Science and Technology of China, Shanghai 201315, China}
\affiliation{$3$Hefei National Laboratory, University of Science and Technology of China, Hefei 230088, China}
\affiliation{$4$New Cornerstone Science Laboratory, School of Emergent Technology, University of Science and Technology of China, Hefei 230026, China}

\date{\today}

\begin{abstract}
    The interference patterns of ultracold atoms, observed after ballistic expansion from optical lattices, encode essential information about strongly correlated lattice systems, including phase coherence and non-local correlations. While the interference of lattice bosons has been extensively investigated, quantitative studies of the lattice fermion interference remain challenging. Here, we report the observation and quantitative characterization of interference patterns in low-temperature, homogeneous fermionic Hubbard gases. We develop a novel method to extract first-order correlations from interference patterns, which directly reflect the short-range phase coherence of lattice fermions. Mapping the nearest-neighbor correlations as a function of lattice filling and interaction strength, we observe a crossover from a metal to a Mott insulator. Moreover, at half filling, the measured correlations agree well with quantum Monte Carlo calculations and remain finite in the regime of strong repulsion, revealing virtual tunneling processes driven by quantum fluctuations.
\end{abstract}

\maketitle

When ultracold atoms in optical lattices are released and undergo ballistic expansion, their spatial coherence manifests as interference patterns in the resulting momentum distribution~\cite{RevModPhys.80.885,greiner2002quantum,PhysRevLett.87.160405}. These patterns reflect the underlying quantum correlations and offer valuable information about the nature of quantum phases in the system~\cite{PhysRevLett.98.080404,bakr2010probing}. For bosonic atoms in optical lattices, the emergence of sharp interference peaks indicates the presence of long-range phase coherence, a hallmark of superfluidity~\cite{greiner2002quantum,greiner2002collapse}. In stark contrast, the interference pattern of a Mott insulator is diffuse and shows no sharp peaks, reflecting the suppression of long-range coherence due to interaction-induced localization~\cite{greiner2002quantum,PhysRevLett.98.080404,bakr2010probing}. Nevertheless, even deep in the Mott insulating state, finite short-range phase coherence persists, which can be attributed to the coherent admixture of virtual particle-hole pairs~\cite{PhysRevLett.95.050404,PhysRevA.72.053606}.

\begin{figure}[tbp]
    \includegraphics[width=1\columnwidth]{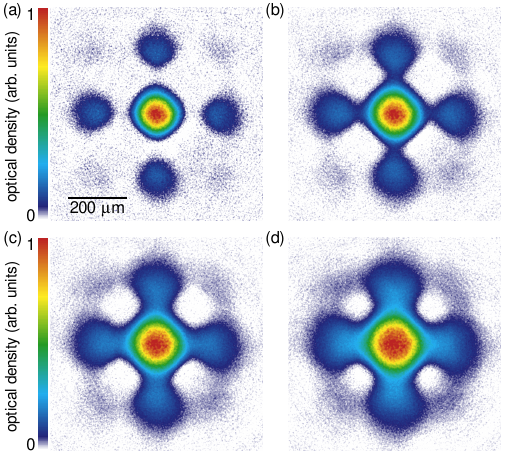}
    \caption{Interference patterns in the non-interacting regime. The lattice filling $n$ in (a)-(d) is 0.25, 0.5, 0.75 and 1, respectively, with a lattice depth of $6.25\ E_\mathrm{r}$. Each image is an average of 100 independent single-shot measurements. \label{fig:Figure 1}}
\end{figure}

For fermionic atoms in optical lattices, the interference patterns encode key physical properties such as quasi-momentum distributions, short-range phase coherence, and first-order correlations~\cite{esslinger2010fermi}. These quantities are crucial for understanding the behaviors of strongly correlated fermions and provide valuable insights into the low-temperature phases of the fermionic Hubbard model (FHM)~\cite{bloch2012quantum,gross2017quantum,Tarruell2018,qin2022hubbard,PhysRevB.111.035123}. Interference patterns have been observed for lattice fermions in both purely fermionic ensembles~\cite{roati2004atom,PhysRevLett.94.080403,chin2006evidence} and Bose-Fermi mixtures~\cite{2015Observation}. Despite these observations, employing them as quantitative probes for exploring the physical properties of strongly correlated systems presents several challenges. First, near half-filling, the quasi-momentum distribution broadens significantly~\cite{PhysRevB.80.075116}, leading to partial overlap of neighboring interference peaks and reduced contrast~\cite{PhysRevLett.94.080403}. Second, at relatively high temperature in previous experiments, thermal broadening of the quasi-momentum distributions blur the interference pattern~\cite{roati2004atom,PhysRevLett.94.080403}, decreasing its visibility. More importantly, inhomogeneity caused by the Gaussian profile of the lattice beams results in spatially varying Hubbard parameters, leading to the coexistence of multiple states across the system~\cite{jordens2008mott,schneider2008metallic}. As a consequence, the observed interference pattern reflects an average over these states, making it difficult to isolate individual contributions and accurately determine the underlying phase coherence and correlations.

In this Letter, we quantitatively characterize the interference pattern of fermions in a homogeneous and low-temperature fermionic Hubbard (FH) system~\cite{PhysRevLett.134.043403,Shao2024antiferromagnetic}, enabling precise extraction of two-dimensional (2D) quasi-momentum distributions over a wide range of lattice fillings up to half-filling. Building on this, we extract first-order correlations $g^{(1)}(d)$ at various distances as functions of lattice filling and interaction strength, providing a detailed characterization of short-range phase coherence in lattice fermions. The diagram of nearest-neighbor correlation shows a clear crossover from a metallic state to a Mott insulator~\cite{PhysRevLett.134.016503}, marked by a gradual suppression of $g^{(1)}(d=1)$ with increasing interactions at half-filling or increasing filling under strong repulsive interactions. At half-filling, the measured $g^{(1)}(d=1)$ is in excellent agreement with quantum Monte Carlo calculations at a temperature of $k_\text{B}T=0.25t$, where $k_\text{B}$ is the Boltzmann constant and $t$ is the hopping rate. In the regime of strong repulsion where fermions are localized, $g^{(1)}(d=1)$ remains finite rather than vanishing, reflecting the presence of virtual particle-hole fluctuations that persist deep in the Mott insulating state~\cite{PhysRevLett.95.050404,endres2011observation}. Our results highlight the rich physics embedded in the interference patterns of lattice fermions, establishing that these patterns can serve as a powerful probe for studying strongly correlated physics.

We perform our experiment with a balanced spin mixture of $^6\mathrm{Li}$ in the two hyperfine states $|1\rangle=|F=1/2,m_F=1/2\rangle$ and $|3\rangle=|F=3/2,m_F=-3/2\rangle$. The atoms are confined in a hybrid trapping potential consisting of a three-dimensional (3D) flat-top optical lattice with a depth of $6.25\ E_\mathrm{r}$ and a cylindrical box trap with an inner diameter of 49.5 $\mu$m and a height of 47.7 $\mu$m~\cite{navon2021quantum,li2022second,2024pseudogap}, providing a spatially uniform lattice potential across approximately $8\times10^5$ lattice sites~\cite{PhysRevLett.134.043403,Shao2024antiferromagnetic}. Here, $E_{\text{r}}/h=h/(8ma^2)=29.30$~kHz denotes the recoil energy, where h is Planck's constant, $m$ represents the mass of $^6\mathrm{Li}$ atom, and $a=532$~nm is the lattice spacing. The lattice filling is controlled by adjusting the final depth of the cylindrical box trap during evaporation, prior to loading the homogeneous Fermi gas into the flat-top optical lattice. The interaction strength $U$ is tuned via a Feshbach resonance by varying the magnetic field, allowing $U/h$ to be controlled over the range $0.00(2)\leq U/h\leq 27.97(4)$~kHz. At half-filling and in the non-interacting regime, the temperature $k_\text{B}T$ is determined to be $0.25(1)t$~\cite{Shao2024antiferromagnetic}, where $t/h=1.40(1)$~kHz is the hopping rate.

\begin{figure}[tbp]
    \includegraphics[width=1\columnwidth]{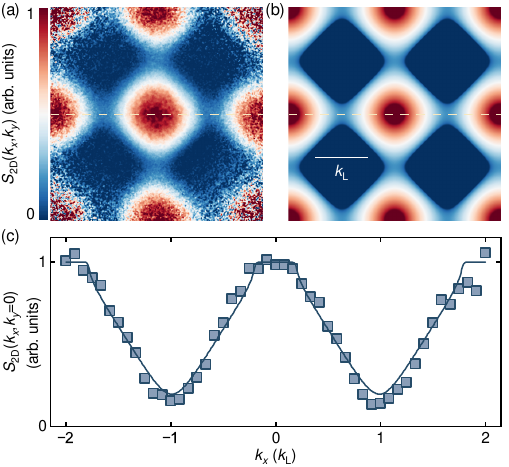}
    \caption{Quasi-momentum distributions of non-interacting homogeneous FH gases at lattice filling $n=0.5$. (a) 2D quasi-momentum distribution in the range from $-2k_\text{L}$ to $2k_\text{L}$, extracted from measured interference pattern shown in Fig.~\ref{fig:Figure 1}(b). (b) Zero-temperature theoretical 2D quasi-momentum distribution. (c) 2D quasi-momentum distribution along the $k_x$-direction at $k_y=0$, indicated by the dashed lines in (a) and (b). The blue squares represent experimental data, and the blue solid line corresponds to zero-temperature theoretical prediction.\label{fig:Figure 2}}
\end{figure}

Once the low-temperature and homogeneous FH gases are prepared, we abruptly turn off all optical trapping potentials, allowing the atoms to expand ballistically for 2~ms. We then measure the resulting density distribution of either spin state $|1\rangle$ or $|3\rangle$ using strong saturation absorption imaging. After time-of-flight, the atomic cloud expands to approximately 15 times its initial size. Consequently, the measured density distribution $\tilde{n}(\boldsymbol{r})$ can be interpreted as the momentum distribution $n(\boldsymbol{k})$~\cite{PhysRevLett.101.155303}, which satisfies:
\begin{equation}
    n(\boldsymbol{k})=|\tilde{w}_0(\boldsymbol{k})|^2S(\boldsymbol{k}).
    \label{Eq1}
\end{equation}
Here, $\tilde{w}_0(\boldsymbol{k})$ is the Fourier transform of the maximally localized Wannier function. The quasi-momentum distribution of fermions in the lattice is given by $S(\boldsymbol{k})=\langle c^\dagger_{\boldsymbol{k}} c_{\boldsymbol{k}}\rangle$, where $ c^\dagger_{\boldsymbol{k}}=\frac{1}{\sqrt{L^3}}\sum_i e^{-i\boldsymbol{k}\cdot\boldsymbol{R}_i} c^\dagger_i$ is the quasi-momentum creation operator with $\boldsymbol{R}_i$ denoting the position of the $i$-th site and $L$ the lattice size. The quasi-momentum distribution is periodic, satisfying $S(\boldsymbol{k} + 2mk_\text{L}\boldsymbol{e_i}) = S(\boldsymbol{k})$, where $m$ is an integer and $\boldsymbol{e}_i$ is the unit vector along the lattice direction. Here, $k_\text{L}=\pi/a$ is the lattice wave vector, corresponding to half the width of the first Brillouin zone. This periodicity leads to the formation of interference patterns in the momentum distributions, which are distinctly observed over a wide range of lattice fillings ($0.25\le n\le 1$), as illustrated in Fig.~\ref{fig:Figure 1}.

Next, we aim to extract the quasi-momentum distribution from the measured interference patterns, a key quantity in strongly correlated systems that is theoretically challenging to compute due to strong interactions. Experimentally, the band mapping technique is commonly used to infer quasi-momentum distributions~\cite{PhysRevLett.94.080403,brown2020angle}. However, unavoidable nonadiabatic interband excitations during lattice ramp-down can introduce systematic deviations, particularly at higher quasi-momenta~\cite{PhysRevA.79.063605,brown2020angle}. In contrast, the measurement of interference patterns is free from this issue. Thus, according to Eq.~(\ref{Eq1}), it not only enables a more accurate determination of the quasi-momentum distribution but also provides a clear delineation of the Brillouin zone boundaries.

In our experiment, the measured momentum distribution is a column-integrated distribution along the $z$-direction, i.e., $n_{2\mathrm{D}}(k_x,k_y) = \int^{+\infty}_{-\infty}n(\boldsymbol{k})\text{d} k_z$. Despite this integration, we find that the measured momentum distribution $n_{\mathrm{2D}}(k_x,k_y)$ retains a factorized form (see Supplementary Material for details):
\begin{equation}
    n_{\mathrm{2D}}(k_x,k_y)= |\tilde{w}_0(k_x,k_y)|^2S_{\mathrm{2D}}(k_x,k_y).
    \label{Eq2}
\end{equation}
Here, $\tilde{w}_0(k_x,k_y)=\tilde{w}_0(k_x)\tilde{w}_0(k_y)$, where $\tilde{w}_0(k_x)$ and $\tilde{w}_0(k_y)$ are the Fourier transforms of the 1D maximally localized Wannier functions in the $x$ and $y$ directions, respectively, which are solely determined by the depths of the flat-top optical lattices. Using intensity modulation spectroscopy~\cite{Shao2024antiferromagnetic}, we calibrate these depths with an uncertainty less than 0.5\%, ensuring a precise determination of $\tilde{w}_0(k_x,k_y)$. Similarly, the integrated quasi-momentum distribution along the $z$-direction is given by $S_{\mathrm{2D}}(k_x,k_y)=\frac{1}{2k_{\text{L}}}\int^{k_\text{L}}_{-k_\text{L}}S(k_x,k_y,k_z)dk_z$, which preserves the periodicity in the $x$-$y$ plane.

Using Eq.~(\ref{Eq2}), we extract the 2D quasi-momentum distribution from the measured interference patterns across various lattice filling and interaction strengths. As an example, Fig.~\ref{fig:Figure 2}(a) presents $S_{\mathrm{2D}}(k_x,k_y)$ at $n=0.5$ in the non-interacting regime, revealing a sharp Fermi surface and a striking periodic structure, which closely align with the zero-temperature theoretical prediction shown in Fig.~\ref{fig:Figure 2}(b). A direct quantitative comparison of 1D quasi-momentum cut in Fig.~\ref{fig:Figure 2}(c) further confirms the excellent agreement between experiment and theory. The Fermi surface and periodic quasi-momentum distribution of non-interacting lattice fermions are fundamental concepts in statistical mechanics, solid-state, and condensed-matter physics. Our experimental results provide vivid and precise demonstrations of these textbook concepts, demonstrating the validity of our approach in obtaining quasi-momentum distributions.

\begin{figure}[tbp]
    \includegraphics[width=1\columnwidth]{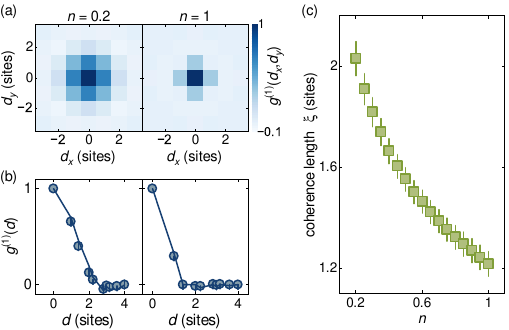}
    \caption{First-order correlations of non-interacting homogeneous FH gases. (a) The first-order correlation function $g^{(1)}(d_x,d_y)$ extracted from 2D quasi-momentum distribution at lattice filling $n=0.2$ (left panel) and $n=1$ (right panel). (b) Azimuthally averaged $g^{(1)}(d)$ as a function of distance $d$ for the two fillings shown in (a). The blue circles show experimental data with error bars representing one standard deviation, and the blue solid lines indicate theoretical predictions at a temperature of $k_BT/t=0.25$. (c) The coherence length $\xi$ as a function of lattice filling. The green squares represent values from Gaussian fits to $g^{(1)}(d)$, with error bars denoting the 95\% confidence intervals. \label{fig:Figure 3}}
\end{figure}

The measurement of the quasi-momentum distribution provides a powerful tool for probing the short-range coherence of atoms in the optical lattice. Qualitatively, coherence is reflected by sharp peaks in the quasi-momentum distribution, while decoherence manifests as a broadening of these peaks or a uniform distribution~\cite{greiner2002quantum,PhysRevLett.98.080404,schneider2008metallic}. These features are directly related to the first-order correlation function, which quantitatively measures the system's coherence. Here, we focus on the normalized first-order correlation function, $g^{(1)}(d) =\langle c^\dagger_d c_{0}\rangle/\sqrt{n_dn_{0}}$, where $d$ denotes the distance between two lattice sites. This correlation is directly related to the quasi-momentum distribution, and can be extracted from $S_{\mathrm{2D}}(k_x,k_y)$ via an inverse Fourier transform, as expressed in the following equation (see Supplementary Material for details):
\begin{equation}
    g^{(1)}(d_x,d_y) = \frac{1}{L^2}\sum_{k_x,k_y}e^{i(k_xd_x+k_yd_y)}\frac{S_{2\mathrm{D}}(k_x,k_y)}{\sqrt{n_{d_x,d_y}n_0}}.
    \label{Eq4}
\end{equation}

Figure~\ref{fig:Figure 3}(a) presents representative first-order correlation functions $g^{(1)}(d_x,d_y)$ for non-interacting, homogeneous FH gases at lattice fillings $n=0.2$ and $n=1$. At low filling ($n=0.2$), correlations extend up to $d=2$, while at half-filling, they decay more rapidly, and no correlations beyond nearest neighbors are resolved within experimental sensitivity. A quantitative comparison between the measured correlations and theoretical predictions at a temperature of $k_BT=0.25t$ shows excellent agreement (see Fig.~\ref{fig:Figure 3}(b)), validating the accuracy of our experimental method. The spatial decay of $g^{(1)}(d)$ is well captured by a Gaussian function, $g^{(1)}(d)=\exp(-2d^2/\xi^2)$, from which we extract the coherence length $\xi$ across different lattice fillings, as shown in Fig.~\ref{fig:Figure 3}(c). In contrast to bosonic superfluids with long-range coherence, fermionic Hubbard gases exhibit only short-range phase coherence, with the coherence length decreasing monotonically as lattice filling increases, from $\xi=2.03(7)$ at $n=0.2$ to $\xi=1.22(5)$ at half-filling. This behavior reflects the intrinsic suppression of coherence in fermionic systems, arising from the Pauli exclusion principle and the gradual filling of low-momentum states. As the filling increases, these states become progressively occupied, limiting coherence to short distances.

\begin{figure}[tbp]
    \includegraphics[width=1\columnwidth]{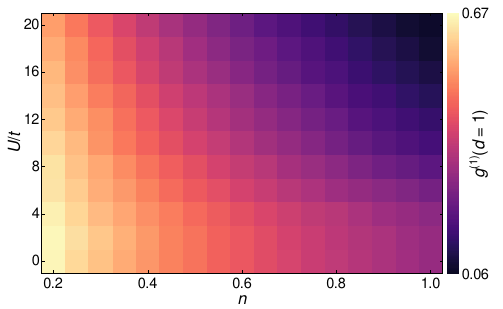}
    \caption{The diagram of $g^{(1)}(d=1)$ as a function of lattice filling and interaction strength. Each data point represents an average of 16 independent measurements.\label{fig:Figure 4}}
\end{figure}

We then measure the diagram of the nearest-neighbor first-order correlation $g^{(1)}(d=1)$ as a function of lattice filling between 0.2 and 1, and interaction strength $U/t$ from 0.00(2) to 20.00(3), as shown in Fig.~\ref{fig:Figure 4}, with the aim of characterizing different states of the FH gases. As the interaction strength increases and the lattice filling approaches half-filling, $g^{(1)}(d=1)$ decreases gradually from a maximum value of 0.659(4) to 0.070(1), dropping by an order of magnitude. This behavior suggests that the system evolves continuously from a metallic state to a Mott insulator, indicating a metal-to-Mott-insulator crossover~\cite{PhysRevLett.134.016503,PhysRevB.111.035123}. Our measurements provide a more comprehensive view of this crossover compared to previous experiments~\cite{jordens2008mott,schneider2008metallic}, such as measuring the doublon fraction as a function of interaction strength at half-filling~\cite{PhysRevLett.134.043403}, which mainly probe the local fermionic density. In contrast, the nearest-neighbor first-order correlation directly captures the short-range spatial correlations across the system. It is sensitive to both the interaction strength and lattice filling, revealing how these variables jointly influence the localization of fermions and drive the crossover process.

\begin{figure}[tbp]
    \includegraphics[width=1\columnwidth]{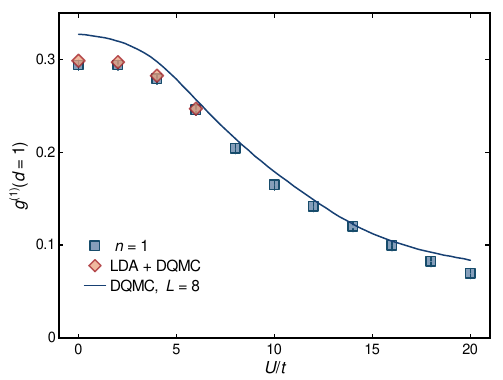}
    \caption{Nearest-neighbor first-order correlations as a function of $U/t$. Blue squares correspond to the experimental $g^{(1)}(d=1)$ at half-filling, while the blue solid line represents the numerical result from DQMC simulations at a temperature of $k_\text{B}T=0.25t$. Red diamonds indicate the LDA results (see text). Each data point is the average of approximately 16 independent measurements, with the error bar denoting the standard deviation. \label{fig:Figure 5}}
\end{figure}

Figure~\ref{fig:Figure 5} further illustrates the measured first-order correlations at half-filling, which is progressively suppressed as the interaction strength increases. Intriguingly, in contrast to a rapid decay to zero upon entering the Mott insulating or antiferromagnetic state, our experimental measurements show that $g^{(1)}(d=1)$ remains finite even at strong interactions, e.g., $U/t=20.00(3)$. This is because, while tunneling is strongly suppressed, short-range correlations can be sustained by quantum fluctuations. More specifically, in the limit of strong interactions, the Hubbard model at half-filling can be approximated by the Heisenberg model~\cite{PhysRev.115.2}, where the superexchange interaction directly arises from second-order virtual tunneling processes between nearest-neighbor lattice sites~\cite{trotzky2008time,2007Controlledexchange}. The superexchange coupling strength, given by $J/t = 4t/U$, is inversely proportional to the interaction strength, leading to a power-law decay of $g^{(1)}(d=1)$ as $U/t$ increases. 

To gain further insights, we compare the experimental results (blue squares) with theoretical predictions (blue curve) obtained from the determinant quantum Monte Carlo (DQMC) method~\cite{PhysRevB.80.075116,PhysRevB.111.035123,SmoQyDQMC}, using a simulation size of $L = 8$ and temperature $k_\text{B}T = 0.25t$. The numerical $g^{(1)}(d=1)$ values as a function of interaction strength agree well with the experimental data for $U/t \gtrsim 6$. However, for $U/t < 6$, the experimental results deviate from the theoretical curve. This is because, near the system boundaries, defined by the repulsive box potential with finite width, the atomic density is suppressed, leading to a deviation from half-filling. This deviation is more pronounced for smaller values of $U$. To address this, we apply the local density approximation (LDA) method, using the DQMC data along with the precise knowledge of the box trap potential, to calculate $g^{(1)}(d=1)$. The theoretical results, shown as red diamonds, almost coincide with the experimental data for $0 \leq U/t \leq 6$.  These results, together with the excellent agreement between theory and experiment, confirm that interference patterns provide a precise method for extracting correlation functions in the FHM.

In summary, we have successfully observed and quantitatively characterized the interference patterns of fermions in a homogeneous optical lattice. By directly extracting the quasi-momentum distribution from these patterns, we have obtained accurate results that perfectly align with theoretical predictions. This advancement has enabled us to measure non-local correlation functions with high precision, offering new insights into the FHM. In the near future, we plan to investigate the non-equilibrium dynamics of fermionic interference, focusing on the collapse and revival of short-range coherence, with the goal of measuring higher-order correlation functions in the FHM~\cite{2015Observation,PhysRevA.90.031602,mahmud2014collapse}. Furthermore, by simultaneously imaging both spin states, we can extract spin-resolved noise correlation functions from the interference patterns~\cite{PhysRevA.70.013603,PhysRevLett.94.110401}. This technique will allow us to directly probe fermion pairing correlations, opening new avenues for exploring exotic phases like the pseudogap~\cite{2024pseudogap} and $d$-wave superconductivity~\cite{RevModPhys.78.17}.

We thank Y. Deng and Q.-J. Chen for discussions. This work is supported by the Innovation Program for Quantum Science and Technology (Grant No. 2021ZD0301900), NSFC of China (Grant No. 11874340), the Chinese Academy of Sciences (CAS), the Anhui Initiative in Quantum Information Technologies, the Shanghai Municipal Science and Technology Major Project (Grant No.2019SHZDZX01), the Basic Science Center Project of NSFC (Grant No. 12488301), and the New Cornerstone Science Foundation.


\clearpage
\onecolumngrid
\appendix
\section*{Interference and short-range correlation in homogeneous fermionic Hubbard gases Supplementary Material}

\section{Section I. Extracting the quasi-momentum distribution}
The momentum distribution of atoms in an optical lattice is given by $n(\boldsymbol{k})=\langle  a^\dagger_{\boldsymbol{k}} a_{\boldsymbol{k}}\rangle$. For simplicity, we omit the spin index both here and in the following discussion. Here, $ a^\dagger_{\boldsymbol{k}}$ is the momentum creation operator, which satisfies the Fourier transform relation with the position operator $ a^\dagger_{\boldsymbol{r}}$ in continuous space:
\begin{equation}
a^\dagger_{\boldsymbol{k}}=\frac{1}{\sqrt{L^3}}\int e^{-i\boldsymbol{k}\cdot\boldsymbol{r}}a^\dagger_{\boldsymbol{r}}\mathrm{d}\boldsymbol{r},
\label{Eq:1}
\end{equation}
where we set $\hbar=1$ and $L$ represents the lattice size. The position operator can be expanded using the creation operator $c^{\dagger}_l$ at the lattice sites as follows:
\begin{equation}
a^{\dagger}_{\boldsymbol{r}}=\sum_l w_0(\boldsymbol{r}-\boldsymbol{R}_l)c^{\dagger}_l.
\label{Eq:2}
\end{equation}
Here, $w_0(\boldsymbol{r}-\boldsymbol{R}_l)$ is the maximally localized Wannier function at the $l$-th lattice site, and $\boldsymbol{R}_l$ is the position of the $l$-th lattice site. Substituting this into Eq.~\ref{Eq:1}, we obtain:
\begin{equation}
\begin{aligned}
    a^{\dagger}_{\boldsymbol{k}}&=\frac{1}{\sqrt{L^3}}\int e^{-i\boldsymbol{k}\cdot\boldsymbol{r}}\sum_l w_0(\boldsymbol{r}-\boldsymbol{R}_l)c^{\dagger}_l\mathrm{d}\boldsymbol{r}\\
    &=\frac{1}{\sqrt{L^3}}\sum_l c^{\dagger}_l\int e^{-i\boldsymbol{k}\cdot\boldsymbol{r}}w_0(\boldsymbol{r}-\boldsymbol{R}_l)\mathrm{d}\boldsymbol{r}\\
    &=\frac{1}{\sqrt{L^3}}\sum_l e^{-i\boldsymbol{k}\cdot\boldsymbol{R}_l}\tilde{w}_0(\boldsymbol{k})c^{\dagger}_l,
\end{aligned}
\label{Eq:3}
\end{equation}
where $\tilde{w}_0(\boldsymbol{k})=\int e^{-i\boldsymbol{k}\cdot\boldsymbol{r}}w_0(\boldsymbol{r})\mathrm{d}\boldsymbol{r}$ is the Fourier transforms of the maximally localized Wannier function $w_0(\boldsymbol{r})$. By substituting this expression for $a^\dagger_{\boldsymbol{k}}$ into the momentum distribution $n(\boldsymbol{k})$, we have:
\begin{equation}
    \begin{aligned}
        n(\boldsymbol{k})&=|\tilde{w}_0(\boldsymbol{k})|^2\frac{1}{L^3}\sum_{l,m}e^{-i\boldsymbol{k}(\boldsymbol{R}_l-\boldsymbol{R}_m)}\langle  c^{\dagger}_l c_m\rangle\\
        &=|\tilde{w}_0(\boldsymbol{k})|^2S(\boldsymbol{k}).
    \end{aligned}
    \label{Eq:4}
\end{equation}
Here, $S(\boldsymbol{k})=\frac{1}{L^3}\sum_{l,m}e^{-i\boldsymbol{k}(\boldsymbol{R}_l-\boldsymbol{R}_m)}\langle  c^{\dagger}_l c_m\rangle=\langle c^{\dagger}_{\boldsymbol{k}}c_{\boldsymbol{k}} \rangle$ is the quasi-momentum distribution, and $c^{\dagger}_{\boldsymbol{k}}=\frac{1}{\sqrt{L^3}}\sum_l e^{-i\boldsymbol{k}\cdot\boldsymbol{R}_l}c^{\dagger}_l$ is the quasi-momentum creation operator. Therefore, the momentum distribution $n(\boldsymbol{k})$ is the product of $|\tilde{w}_0(\boldsymbol{k})|^2$ and quasi-momentum distribution $S(\boldsymbol{k})$.

In our experiment, the momentum distribution measured is the column-integrated momentum distribution along the imaging $z$-direction, given by:
\begin{equation}
    \begin{aligned}
            n_{\mathrm{2D}}(k_x,k_y)&=\int^{+\infty}_{-\infty}n(k_x,k_y,k_z)\mathrm{d}k_z\\
            &=\int^{+\infty}_{-\infty}|\tilde{w}_0(k_x,k_y,k_z)|^2S(k_x,k_y,k_z)\mathrm{d}k_z\\
            &= |\tilde{w}_0(k_x)\tilde{w}_0(k_y)|^2\int^{+\infty}_{-\infty}|\tilde{w}_0(k_z)|^2S(k_x,k_y,k_z)\mathrm{d}k_z,
    \end{aligned}
    \label{Eq:5}
\end{equation}
where $\tilde{w}_0(k_x,k_y,k_z)$ can be decomposed into the product of $\tilde{w}_0(k_x)$, $\tilde{w}_0(k_y)$ and $\tilde{w}_0(k_z)$ in the three independent directions. Solving the Schrödinger equation for the periodic lattice potential with $L$ sites, the on-site Wannier wave function can be expressed as:
\begin{equation}
    w_0(z) = \frac1{\sqrt L}\sum_{q,m}C_{qm}e^{i(q+2mk_\text{L})z}.
    \label{Eq:6}
\end{equation}
Here, the sum over $q$ runs within the first Brillouin zone, $-k_\text{L}\le q<k_\text{L}$, where $k_\text{L}=\pi/a$ is the magnitude of lattice wave vector and $a$ is the lattice spacing. The index $m$ is an integer, and the coefficients $C_{qm}$ satisfy $\sum_{m\in\mathcal{Z}}|C_{qm}|^2=1$. By performing a Fourier transform on the Wannier function at the lattice sites, we obtain its momentum-space representation:
\begin{equation}
    \tilde{w}_0(k_z) = \frac1{\sqrt L}\sum_{q,m}C_{qm}\delta(k_z-q-2mk_\text{L}).
    \label{Eq:7}
\end{equation}
Substituting this expression for $\tilde{w}_0(k_z)$ into Eq.~\ref{Eq:5} for $n_{\mathrm{2D}}(k_x,k_y)$ and utilizing the periodicity of the quasi-momentum distribution, $S(k_x,k_y,k_z+2mk_\text{L}) = S(k_x,k_y,k_z)$ with $m\in\mathcal{Z}$, we obtain:
\begin{equation}
    \begin{aligned}
        \int^{+\infty}_{-\infty}|\tilde{w}_0(k_z)|^2S(k_x,k_y,k_z)\mathrm{d}k_z&=\int^{+\infty}_{-\infty}\frac1L\sum_{q,m}\sum_{p,n}C_{qm}C_{pn}\delta(k_z-q-2mk_\text{L})\delta(k_z-p-2nk_\text{L})S(k_x,k_y,k_z)dk_z\\
        &=\frac1L\sum_{q,m}|C_{qm}|^2S(k_x,k_y,q+2mk_\text{L})=\frac1L\sum_{q,m}|C_{qm}|^2S(k_x,k_y,q)\\
        &=\frac1L\sum_qS(k_x,k_y,q)=\frac{1}{2k_\text{L}}\int^{k_\text{L}}_{-k_\text{L}}S(k_x,k_y,k_z)\mathrm{d}k_z = S_{\mathrm{2D}}(k_x,k_y).
    \end{aligned}
    \label{Eq:8}
\end{equation}
Therefore, the column-integrated momentum distribution can also be expressed as the product of $|\tilde{w}_0(k_x)\tilde{w}_0(k_y)|^2$ and the 2D quasi-momentum distribution:
\begin{equation}
    n_{\mathrm{2D}}(k_x,k_y)=|\tilde{w}_0(k_x)\tilde{w}_0(k_y)|^2\times S_{\mathrm{2D}}(k_x,k_y).
    \label{Eq:9}
\end{equation}

\section{Section II. Obtaining the first-order correlation}
In a three-dimensional optical lattice, the correlation function depends only on the relative distance $\boldsymbol{d}$ between lattice sites, owing to the discrete translational symmetry of the system. The normalized first-order correlation, defined as $g^{(1)}(d) =\langle c^\dagger_d c_0\rangle/\sqrt{n_dn_0}$, is closely related to the 2D quasi-momentum distribution $S_{\mathrm{2D}}(k_x,k_y)$:
\begin{equation}    
    \begin{aligned}
    	S_{\mathrm{2D}}(k_x,k_y)&= \frac{1}{2k_\text{L}}\int^{ k_\text{L}}_{- k_\text{L}}S_{\mathrm{3D}}(\boldsymbol{k})\mathrm{d}k_z=\frac{1}{2k_\text{L}}\int^{ k_\text{L}}_{- k_\text{L}}\frac{1}{L^3}\sum_{lm}e^{-i\boldsymbol{k}(\boldsymbol{R}_l-\boldsymbol{R}_m)}\langle \hat{c}^\dagger_l\hat{c}_m\rangle\mathrm{d}k_z\\
        &=\frac{1}{2k_\text{L}}\int^{ k_\text{L}}_{- k_\text{L}}\sum_{\boldsymbol{d}}e^{-i\boldsymbol{k}\cdot\boldsymbol{d}}\langle  c^{\dagger}_{\boldsymbol{d}} c_0\rangle\mathrm{d}k_z=\frac{1}{2k_\text{L}}\sum_{\boldsymbol{d}}e^{-i(k_xd_x+k_yd_y)}\langle c^{\dagger}_{\boldsymbol{d}} c_0 \rangle\int^{ k_\text{L}}_{- k_\text{L}}e^{-ik_zd_z}\mathrm{d}k_z\\
        &=\sum_{\boldsymbol{d}}\delta(d_z)e^{-i(k_xd_x+k_yd_y)}\langle c^{\dagger}_{\boldsymbol{d}} c_0 \rangle=\sum_{d_x,d_y}e^{-i(k_xd_x+k_yd_y)}\langle c^{\dagger}_{d_x,d_y} c_0 \rangle
    \end{aligned}
    \label{Eq:10}
\end{equation}
where the $\delta(d_z)$ are the $\delta$-functions. Thus, by performing an inverse Fourier transform, the first-order correlation $g^{(1)}(d_x,d_y)$ can be obtained from $S_{\mathrm{2D}}(k_x,k_y)$:
\begin{equation}
    g^{(1)}(d_x,d_y) =\frac{\langle c^\dagger_{d_x,d_y} c_0 \rangle}{\sqrt{n_{d_x,d_y} n_0}} = \frac{1}{L^2}\sum_{k_x,k_y}e^{i(k_xd_x+k_yd_y)}\frac{S_{2\mathrm{D}}(k_x,k_y)}{\sqrt{n_{d_x,d_y}n_0}}
    \label{Eq:13}
\end{equation}

\end{document}